\pdfoutput=1

\documentclass[aps,prl,twocolumn,groupedaddress,showpacs]{revtex4}

\usepackage{amssymb}
\usepackage{amsmath}
\usepackage{graphicx}

\begin{document}

\title{Charge and spin state readout of a double quantum dot coupled to a resonator}
\author{K. D. Petersson}
\email{kpeterss@princeton.edu}
\altaffiliation{Present address: Department of Physics, Princeton University, Princeton, New Jersey, 08544, USA}
\author{C. G. Smith}
\email{cgs4@cam.ac.uk}
\author{D. Anderson}
\author{P. Atkinson}
\altaffiliation{Present address: IFW Dresden, Helmholtzstra\ss{}e 20, 01069 Dresden, Germany.}
\author{G. A. C. Jones}
\author{D. A. Ritchie}
 \affiliation{Cavendish Laboratory, JJ Thomson Road, Cambridge CB3 0HE, United Kingdom}
 
\date{\today}


\begin{abstract}
State readout is a key requirement for a quantum computer. For semiconductor-based qubit devices it is usually accomplished using a separate mesoscopic electrometer. Here we demonstrate a simple detection scheme in which a radio-frequency resonant circuit coupled to a semiconductor double quantum dot is used to probe its charge and spin states. These results demonstrate a new non-invasive technique for measuring charge and spin states in quantum dot systems without requiring a separate mesoscopic detector.
\end{abstract}


\maketitle

State readout is a key requirement for a quantum information processor. For semiconductor-based qubit devices this is usually accomplished using a separate mesoscopic electrometer such as a quantum point contact (QPC) \cite{field93}. Non-invasive detection using a QPC has been critical to recent advances in coherent control of few-electron double quantum dots \cite{petta05}, allowing their charge configurations to be mapped in regimes where transport through the double dot itself is not measurable \cite{elzerman03}. Furthermore, through spin-to-charge conversion techniques, spin state measurements of single and double dots have been demonstrated using QPCs \cite{elzerman04, johnson05}.

Resonant microwave circuits have played an important role in performing sensitive measurements of mesoscopic devices. In the case of superconducting charge qubits, earlier schemes used a proximal single electron transistor (SET) charge detector to perform state readout \cite{lehnert03, duty04}. This SET was coupled to an impedance matching rf resonant circuit to form an rf-SET which had dramatically improved sensitivity and bandwidth \cite{schoelkopf98}. Similarly, an rf-QPC has also been realised for broadband charge detection of semiconductor quantum dot devices \cite{cassidy07, reilly07}. 

More recently, state readout of a superconducting charge qubit has been accomplished by directly coupling it to a microwave resonant circuit \cite{wallraff04, duty05, sillanpaa05}. Working in the dispersive regime where the resonator and qubit bandgap energies are detuned, the qubit has a state-dependent `quantum capacitance' which causes a shift in the resonator frequency \cite{blais04}. This frequency shift can then be detected using standard homodyne detection techniques. 

In this Letter, we demonstrate dispersive detection using a resonant circuit coupled directly to an AlGaAs/GaAs few-electron double quantum dot device. This allows us to probe the charge state of a single electron confined to a double quantum dot. We also perform a spin-state measurement for a pair of electrons using the resonant circuit. These results demonstrate a new and non-invasive technique for measuring charge and spin states in semiconductor quantum dot systems without the need for a separate mesoscopic detector.

\begin{figure}
\begin{center}
		\includegraphics[scale=1]{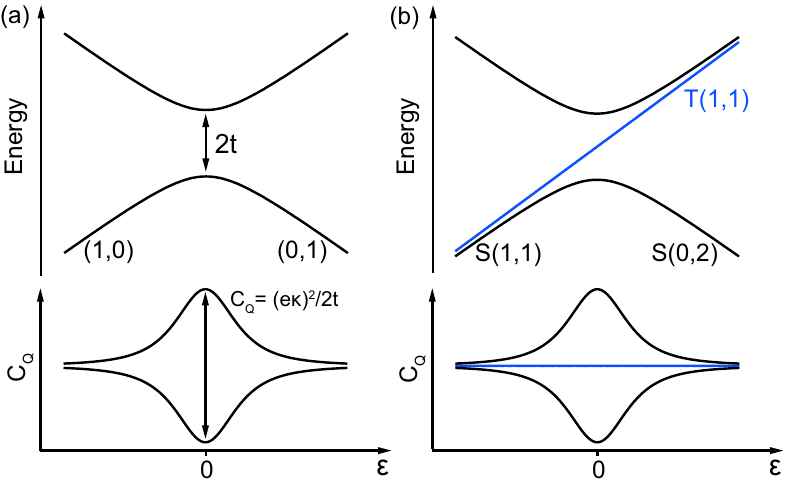}
\caption{\label{fig1} Energy band diagram and quantum capacitance as a function of level detuning $\epsilon$ for a double quantum dot with (a) one electron and (b) two electrons. At large detunings, $\left(m, n \right)$ denotes the electron occupancy of the left and right dots and in (b) the singlet and triplet spin configurations are indicated by $S$ and $T$ respectively \cite{plane}. In (b) the T(1,1)-T(0,2) transition occurs at large positive detuning and is not shown.}
\end{center}	
\end{figure}

In using the resonator to probe the state of a double quantum dot we consider the double dot as a charge qubit where a single electron occupies either the ground state of one dot or the other \cite{hayashi03}. The Hamiltonian for this two level system is given by $H = \frac{1}{2}\epsilon\sigma^z+t\sigma^x$, where $\epsilon$ is the detuning between the two dot chemical potential energies and $t$ is the interdot tunnel coupling energy which mixes the discrete charge states. The ground and excited state energies for our qubit are given by $E_{\pm} = \pm \frac{1}{2} \sqrt{\epsilon^2 + \left(2t\right)^2}$. The bandstructure for a charge qubit is shown in Fig. 1(a). The quantum capacitance of the system is given by,
\begin{equation} \label{curve} C_Q^{\pm} = -\left(e\kappa\right)^2 \frac{ \partial^2 E_{\pm}}{\partial^2 \epsilon}.\end{equation}
Here $\kappa \leq 1$ is used to convert between the resonator's voltage $V_{D1}$ and detuning energy $\epsilon$ ($\Delta \epsilon = -e\kappa \Delta V_{D1}$) and determines the coupling between the resonator and the double dot. Equation \eqref{curve} demonstrates how the quantum capacitance of the qubit is a function of its band curvature or polarizability. This is in contrast to a charge detector which effectively measures the slope of the qubit's energy splitting. As shown in Fig. 1(a), extrema in curvature occur at the interdot charge transition ($\epsilon = 0$) where the qubit is most readily polarizable with the quantum capacitance given by,
\begin{equation}C_Q^{\pm}\left(\epsilon = 0 \right)  = \mp \frac{1}{2}\frac{\left(e \kappa\right)^2}{2t}.\end{equation}
With $\kappa = 1$ and $2t = 1$ GHz, this gives $C_Q^{\pm}\left(\epsilon = 0 \right) \approx \mp 19$ fF. The quantum capacitance of a double quantum dot has been very recently observed \cite{ota09}. In these experiments, however, the measurement still relied upon a QPC charge detector coupled to the double dot.

In the case of a two-electron double dot, exchange coupling results in the singlet and three triplet states being polarizable at different detunings, separated by the relatively large ($\sim 300$ $\mu$eV) singlet-triplet energy spacing. The band structure for this case at zero magnetic field is shown in Fig. 1(b). At zero detuning the quantum capacitance of the ground singlet state is at a maximum whereas it remains zero for the triplet states, thereby allowing these different spin configurations to be distinguished.

\begin{figure}
\begin{center}
		\includegraphics[scale=1]{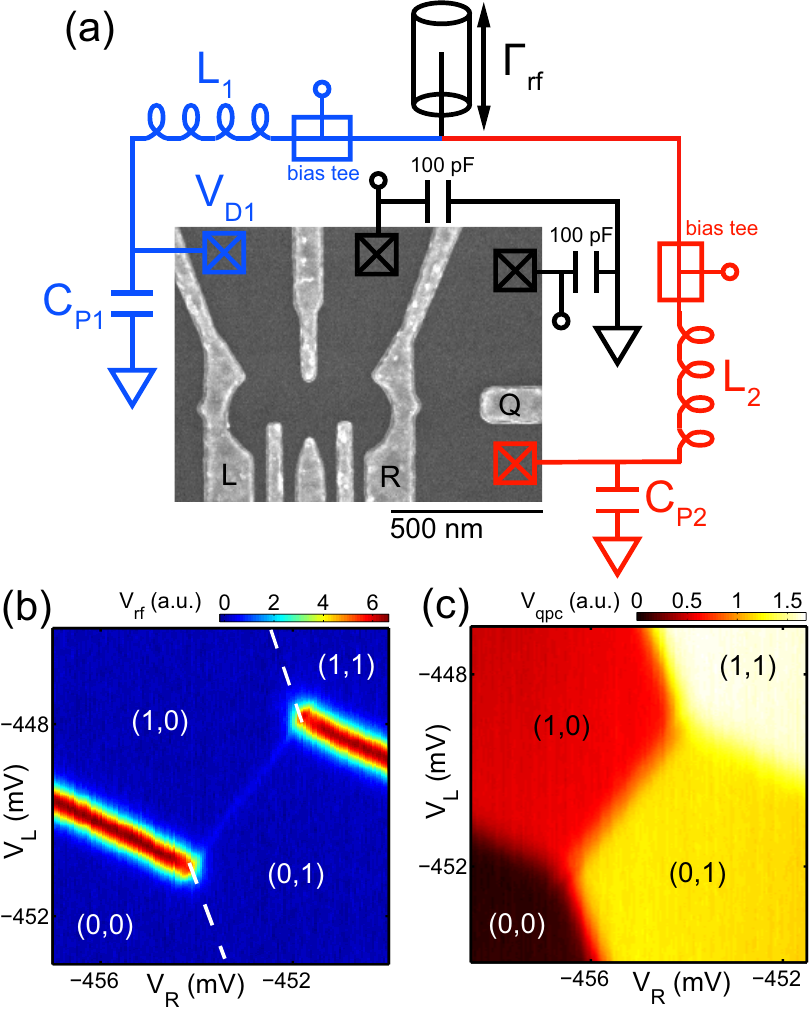}
\caption{\label{fig2}  (a) Schematic of resonator circuit and scanning electron micrograph of double quantum dot device similar to the one measured. The double quantum dot is defined using Ti/Au gates on a GaAs/AlGaAs heterostructure with a two dimensional electron gas (2DEG) 90nm below the surface. Measurements are made in a dilution refrigerator with an effective base electron temperature of $\approx 185$ mK, as determined from thermal broadening of the interdot charge transition \cite{dicarlo04}. Attenuated coax lines are coupled to the left and right gates via bias tees at the mixing chamber. (b) Demodulated response of the resonant circuit as a function of the left and right gates voltages of the double quantum dot. Absolute electron numbers for the left and right dots are indicated in brackets. (c) rf-QPC response for the same gate voltage settings as in (b).  }
\end{center}	
\end{figure}
To detect the quantum capacitance signal we couple the double quantum dot to a resonant circuit which we measure using rf reflectometry \cite{schoelkopf98}. A schematic of the measurement circuit is shown in Fig. 2(a), along with a scanning electron micrograph of a double quantum dot device similar to the one measured. The LC resonant circuit, formed using a chip inductor $L_1 = 330$ nH and its parasitic capacitance to ground $C_{P1}$, is connected to the left reservoir of the double dot device. The resonator is probed using an rf carrier signal at around the resonant frequency, $f_1 \approx 385$ MHz. The reflected signal is amplified using a low-noise cryogenic amplifier and then, at room temperature, further amplified, demodulated and filtered for sampling. The demodulated response is sensitive to both the phase and amplitude of the reflected carrier. The quantum capacitance of the double quantum dot modifies the total resonator capacitance $C_{P1}$ which, in turn, modifies the phase of the reflected carrier.

Using frequency multiplexing \cite{buehler04}, a second resonant circuit, operating at around $f_2 \approx 326$ MHz ($L_2 = 470$ nH) is coupled to an adjacent quantum point contact such that it is configured as an rf-QPC for fast charge detection. We address either the double quantum dot resonator or the rf-QPC by switching the carrier frequency. 

\begin{figure}
\begin{center}
		\includegraphics[scale=1]{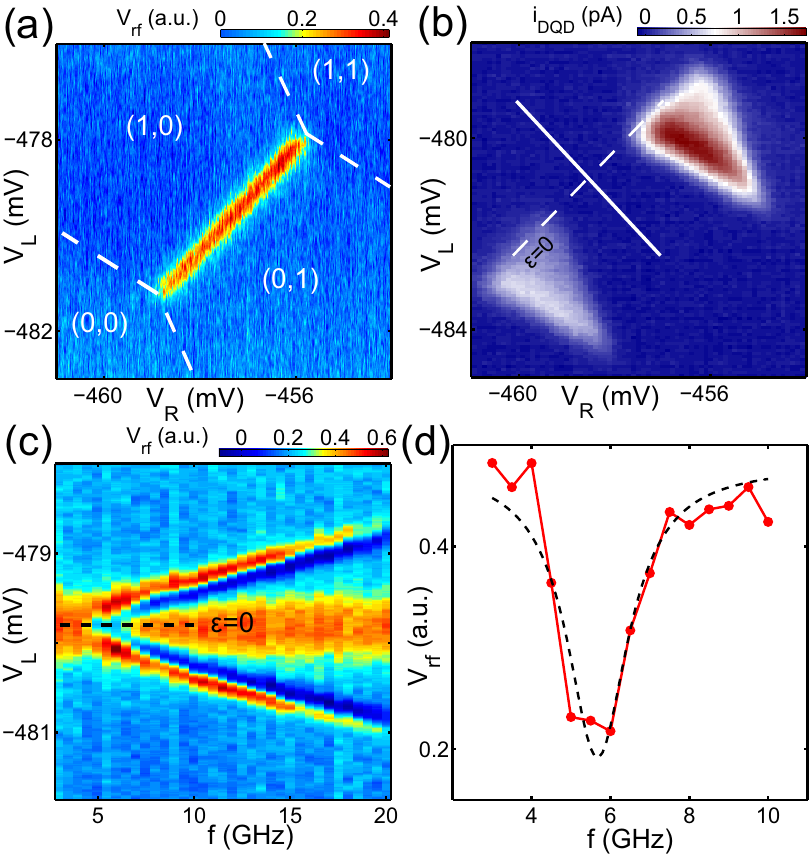}
\caption{\label{fig3} (a) Demodulated response of the resonant circuit as a function of the left and right gate voltages of the double quantum dot. In this case, tunnelling between the left dot and its reservoir is strongly suppressed. (b) Measured current through the device at the same gate tuning with a $V_{D1} \approx 300$ $\mu$V bias applied. (c) Microwave spectroscopy data showing the demodulated resonator response as $V_L$ is swept across the interdot charge transition and the microwave frequency is stepped. (d) Spectroscopy data at $\epsilon =0$ [along the dashed line in (c)] which we fit to a Lorentzian [dashed line].}
\end{center}	
\end{figure}
Figure 2(b) shows the demodulated resonator response as a function of the left and right gate voltages around the $(1, 0)/(0, 1)$ charge transition. The corresponding rf-QPC detector signal is shown in Fig. 2(c). Where the electrochemical potential of the left dot aligns with the Fermi level of the left lead we observe a strong change in the resonator signal as electrons tunnel on and off the dot in response to the carrier drive. This is despite current through the device being blocked along this charge transition (and away from the interdot transition). A similar `capacitance mode' response has previously been observed for a single quantum dot device where tunnelling through one of its two leads is strongly suppressed \cite{cheong02}.

In Fig. 2(b) there is also a weak response along the interdot charge transition. To more clearly observe this signal we adjust the tuning of the device such that tunnelling between the left dot and its lead is strongly suppressed. Figure 3(a) maps the resonator response in this regime. The carrier excitation power is $\sim -93$ dBm ($V_{D1} \sim 120$ $\mu$V) and the rf-QPC is pinched-off. The maximum change in the resonator response occurs at zero detuning and is attributable to quantum capacitance of the double dot. Figure 3(b) shows the current through the device in this same regime with a $300$ $\mu$V bias applied. Along the solid line shown in Fig. 3(b) the current through the device is not measurable ($\lesssim 100$ fA, corresponding to a tunnel rate less than $\sim 1$ MHz), highlighting the non-invasive nature of the resonant circuit measurement.

To distinguish the quantum capacitance of the ground and excited charge states we apply microwaves to drive the double dot into its excited state. In Fig. 3(c) we perform microwave spectroscopy by sweeping across $\epsilon = 0$ and stepping the frequency of a microwave source which is coupled to gate $R$. Figure 3(d) plots the resonator response as a function of frequency at zero detuning [along the dashed line in Fig. 3(c)]. Where the microwave frequency matches the band gap energy $2t$ we observe a dip in the resonator response indicative that the qubit is being resonantly driven to its excited state. Fitting the dip to a Lorentzian, from the half-width at half-maximum, we extract a coherence time of $T_2 \approx 1.1$ ns, consistent with measurements of the decay time for coherent oscillations of a semiconductor charge qubit \cite{hayashi03}. 

We next consider how the resonator can be used for spin state readout. We first perform a spin state measurement using the rf-QPC \cite{petta05}. Using the pulse sequence shown in Fig. 4(a) with $B=0$ \cite{field} we first prepare the double dot in the $S(0,2)$ singlet state by pulsing out to point $P_A$ for 200 ns and then returning to $P_B$ for 200 ns. The two electrons are then separated by moving to point $S$ deep in the $(1, 1)$ charge state for time $\tau_s$. During this separation time, the differences in the effective fields of the nuclei for the two dots mixes the singlet state with the triplet states. Returning to point $M$ inside the dashed region for $T_M \sim 1.5$ $\mu$s gives a measurement of the spin state as transitions to the (0,2) charge configuration are blocked for the triplet states. Averaging the rf-QPC signal over many repetitions of the pulse sequence and different configurations of the fluctuating nuclear field gradient then gives an ensemble measurement of the return probability. As shown in Fig. 4(b), repeating this measurement for different separation times $\tau_s$ we observe a decay in the return probability on a characteristic time scale given by the inhomogeneous spin coherence time, $T_2^*$. We fit our data using a Gaussian decay \cite{churchill09} with the probability of returning to the (0,2) charge state given by $P_S = 1 - C\left(1- e^{-\left(\tau_s/T_2^*\right)^2}\right)$. A least squares fit gives $T_2^* = 5.6$ ns and $C = 0.60$, consistent with other measurements \cite{petta05, reilly08}.

To perform a spin state measurement using the resonator, we repeat the same pulse sequence as before, however, this time we measure the response with point $M$ at zero detuning as indicated in Fig. 4(c). As we are unable to polarize the double dot if it is in one of the triplet states we observe a decay in the resonator response as a function of $\tau_s$. In Fig. 4(d) we fit the normalised peak height $V_{rf}\left( \tau_s\right)/V_{rf}\left( 0\right)$ to a Gaussian and extract $T_2^* = 5.5$ ns and $C = 0.63$ consistent with our measurement using the rf-QPC. As expected, repeating this same pulse sequence at the (1,0)/(0,1) transition we do not observe any such decay.

\begin{figure}
\begin{center}
		\includegraphics[scale=1]{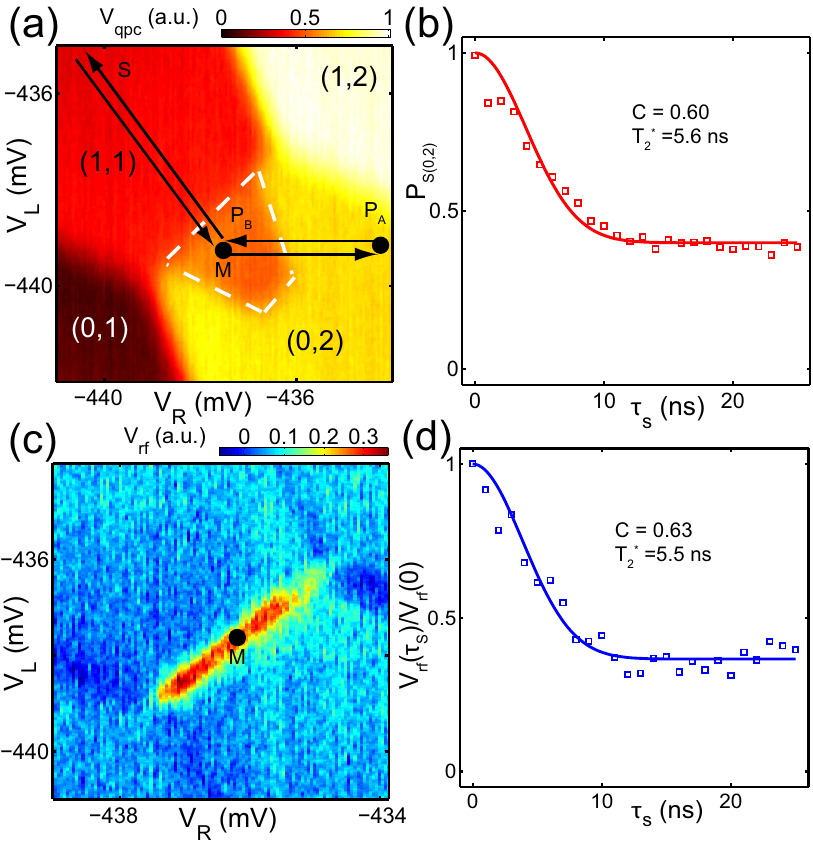}
\caption{\label{fig4} (a) rf-QPC response at the (1,1)/(0,2) interdot charge transition with the pulse sequence indicated by the arrows applied ($\tau_s = 25$ ns). (b) Singlet return probability as a function of $\tau_s$. (c) Resonator response at the (1,1)/(0,2) interdot charge transition (with no pulses applied). (d) Normalised decay in the resonator response at the point $M$ in (c) as a function of $\tau_s$ using the pulse sequence shown in (a). The data points in (b) and (d) are averaged over three data sets.}
\end{center}	
\end{figure}
From the data in Fig. 3(a) which were acquired with an effective bandwidth of $B =3$ Hz, (3 kHz low-pass filter and $10^3$ averages) we estimate a minimum detection time of $\tau_m \sim 4$ ms for a signal-to-noise ratio of one. This measured minimum detection time is comparable with typical spin lifetimes in GaAs \cite{elzerman04}. We expect that this time could be improved by a factor of $\sim 30$ with optimal driving and by reducing resonator losses. We also expect that our estimate of the detection time is limited by $1/f$ noise. Following Johansson et al. \cite{johansson06}, a lower limit for the minimum detection time is given by,
\begin{equation}
\tau_m^{min} \approx \dfrac{k_B T_N C_{P1}^2 Z_0}{8 \beta^2 e^2 \kappa^2},
\end{equation}
where $T_N$ is the system noise temperature, $Z_0 = 50$ $\Omega$, $\beta = \sqrt{2^{2/3}-1}/2$. For our experiment we estimate $T_N \sim 7$ K and $\kappa \sim 0.14$, giving $\tau_m^{min} \sim 2$ $\mu$s. 

Significant improvements in $\tau_m$ could be made by increasing the coupling $\kappa$ between the resonator and double dot. In the case of gated lateral AlGaAs/GaAs quantum dots this is difficult as the dots are formed relatively far from the surface and the total device capacitance is divided amongst the leads and the several device gates. However, it should be feasible to engineer much stronger couplings in nanowire or nanotube devices where there is already natural confinement in an additional dimension. In the case of nanotube devices where the leads can dominate the capacitance \cite{andresen08}, we expect that the coupling could be made $\kappa \gtrsim 0.5$.

In summary, we have demonstrated state detection for a semiconductor double quantum dot device using a resonant circuit. We have shown how this technique can be used to measure coherence times in quantum dot devices. This scheme has several advantages over charge detection using an rf-QPC or rf-SET. Firstly, unlike a QPC which generates broadband shot noise which can result in undesired transitions \cite{gustavsson07}, back-action noise of the resonator is at a well-defined frequency and coupling to the electrical environment can be carefully engineered to suppress charge relaxation. Secondly, dispersive readout permits qubit operation at zero detuning where the gradient of the charge qubit splitting is zero and hence dephasing due to charge noise is minimal. Finally, dispersive readout can greatly simplify device fabrication as a secondary mesoscopic charge detector is no longer required. This is particuarly relevant for complex multidot devices \cite{petersson09}, or nanowire \cite{hu07} or carbon nanotube devices \cite{biercuk06, churchill09}, where yields are typically low and integrating a charge detector is non-trivial. 

\begin{acknowledgments}
We thank A. J. Ferguson and T. Duty for invaluable discussions and S. J. Chorley, J. C. Frake and A. Corcoles for technical contributions. This work was supported by the EPSRC.
\end{acknowledgments}



\begin{thebibliography}{29}
\expandafter\ifx\csname natexlab\endcsname\relax\def\natexlab#1{#1}\fi
\expandafter\ifx\csname bibnamefont\endcsname\relax
  \def\bibnamefont#1{#1}\fi
\expandafter\ifx\csname bibfnamefont\endcsname\relax
  \def\bibfnamefont#1{#1}\fi
\expandafter\ifx\csname citenamefont\endcsname\relax
  \def\citenamefont#1{#1}\fi
\expandafter\ifx\csname url\endcsname\relax
  \def\url#1{\texttt{#1}}\fi
\expandafter\ifx\csname urlprefix\endcsname\relax\def\urlprefix{URL }\fi
\providecommand{\bibinfo}[2]{#2}
\providecommand{\eprint}[2][]{\url{#2}}

\bibitem[{\citenamefont{Field et~al.}(1993)\citenamefont{Field, Smith, Pepper,
  Ritchie, Frost, Jones, and Hasko}}]{field93}
\bibinfo{author}{\bibfnamefont{M.}~\bibnamefont{Field}},
  \bibinfo{author}{\bibfnamefont{C.~G.} \bibnamefont{Smith}},
  \bibinfo{author}{\bibfnamefont{M.}~\bibnamefont{Pepper}},
  \bibinfo{author}{\bibfnamefont{D.~A.} \bibnamefont{Ritchie}},
  \bibinfo{author}{\bibfnamefont{J.~E.~F.} \bibnamefont{Frost}},
  \bibinfo{author}{\bibfnamefont{G.~A.~C.} \bibnamefont{Jones}},
  \bibnamefont{and} \bibinfo{author}{\bibfnamefont{D.~G.} \bibnamefont{Hasko}},
  \bibinfo{journal}{Phys. Rev. Lett.} \textbf{\bibinfo{volume}{70}},
  \bibinfo{pages}{1311} (\bibinfo{year}{1993}).

\bibitem[{\citenamefont{Petta et~al.}(2005)\citenamefont{Petta, Johnson,
  Taylor, Laird, Yacoby, Lukin, Marcus, Hanson, and Gossard}}]{petta05}
\bibinfo{author}{\bibfnamefont{J.~R.} \bibnamefont{Petta}},
  \bibinfo{author}{\bibfnamefont{A.~C.} \bibnamefont{Johnson}},
  \bibinfo{author}{\bibfnamefont{J.~M.} \bibnamefont{Taylor}},
  \bibinfo{author}{\bibfnamefont{E.~A.} \bibnamefont{Laird}},
  \bibinfo{author}{\bibfnamefont{A.}~\bibnamefont{Yacoby}},
  \bibinfo{author}{\bibfnamefont{M.~D.} \bibnamefont{Lukin}},
  \bibinfo{author}{\bibfnamefont{C.~M.} \bibnamefont{Marcus}},
  \bibinfo{author}{\bibfnamefont{M.~P.} \bibnamefont{Hanson}},
  \bibnamefont{and} \bibinfo{author}{\bibfnamefont{A.~C.}
  \bibnamefont{Gossard}}, \bibinfo{journal}{Science}
  \textbf{\bibinfo{volume}{309}}, \bibinfo{pages}{2180} (\bibinfo{year}{2005}).

\bibitem[{\citenamefont{Elzerman et~al.}(2003)\citenamefont{Elzerman, Hanson,
  Greidanus, Willems~van Beveren, De~Franceschi, Vandersypen, Tarucha, and
  Kouwenhoven}}]{elzerman03}
\bibinfo{author}{\bibfnamefont{J.~M.} \bibnamefont{Elzerman}},
  \bibinfo{author}{\bibfnamefont{R.}~\bibnamefont{Hanson}},
  \bibinfo{author}{\bibfnamefont{J.~S.} \bibnamefont{Greidanus}},
  \bibinfo{author}{\bibfnamefont{L.~H.} \bibnamefont{Willems~van Beveren}},
  \bibinfo{author}{\bibfnamefont{S.}~\bibnamefont{De~Franceschi}},
  \bibinfo{author}{\bibfnamefont{L.~M.~K.} \bibnamefont{Vandersypen}},
  \bibinfo{author}{\bibfnamefont{S.}~\bibnamefont{Tarucha}}, \bibnamefont{and}
  \bibinfo{author}{\bibfnamefont{L.~P.} \bibnamefont{Kouwenhoven}},
  \bibinfo{journal}{Phys. Rev. B} \textbf{\bibinfo{volume}{67}},
  \bibinfo{pages}{161308} (\bibinfo{year}{2003}).

\bibitem[{\citenamefont{Elzerman et~al.}(2004)\citenamefont{Elzerman, Hanson,
  Willems~van Beveren, Witkamp, Vandersypen, and Kouwenhoven}}]{elzerman04}
\bibinfo{author}{\bibfnamefont{J.~M.} \bibnamefont{Elzerman}},
  \bibinfo{author}{\bibfnamefont{R.}~\bibnamefont{Hanson}},
  \bibinfo{author}{\bibfnamefont{L.~H.} \bibnamefont{Willems~van Beveren}},
  \bibinfo{author}{\bibfnamefont{B.}~\bibnamefont{Witkamp}},
  \bibinfo{author}{\bibfnamefont{L.~M.~K.} \bibnamefont{Vandersypen}},
  \bibnamefont{and} \bibinfo{author}{\bibfnamefont{L.~P.}
  \bibnamefont{Kouwenhoven}}, \bibinfo{journal}{Nature}
  \textbf{\bibinfo{volume}{430}}, \bibinfo{pages}{431} (\bibinfo{year}{2004}).

\bibitem[{\citenamefont{Johnson et~al.}(2005)\citenamefont{Johnson, Petta,
  Taylor, Yacoby, Lukin, Marcus, Hanson, and Gossard}}]{johnson05}
\bibinfo{author}{\bibfnamefont{A.~C.} \bibnamefont{Johnson}},
  \bibinfo{author}{\bibfnamefont{J.~R.} \bibnamefont{Petta}},
  \bibinfo{author}{\bibfnamefont{J.~M.} \bibnamefont{Taylor}},
  \bibinfo{author}{\bibfnamefont{A.}~\bibnamefont{Yacoby}},
  \bibinfo{author}{\bibfnamefont{M.~D.} \bibnamefont{Lukin}},
  \bibinfo{author}{\bibfnamefont{C.~M.} \bibnamefont{Marcus}},
  \bibinfo{author}{\bibfnamefont{M.~P.} \bibnamefont{Hanson}},
  \bibnamefont{and} \bibinfo{author}{\bibfnamefont{A.~C.}
  \bibnamefont{Gossard}}, \bibinfo{journal}{Nature}
  \textbf{\bibinfo{volume}{435}}, \bibinfo{pages}{925} (\bibinfo{year}{2005}),
  ISSN \bibinfo{issn}{0028-0836}.

\bibitem[{\citenamefont{Lehnert et~al.}(2003)\citenamefont{Lehnert, Bladh,
  Spietz, Gunnarsson, Schuster, Delsing, and Schoelkopf}}]{lehnert03}
\bibinfo{author}{\bibfnamefont{K.~W.} \bibnamefont{Lehnert}},
  \bibinfo{author}{\bibfnamefont{K.}~\bibnamefont{Bladh}},
  \bibinfo{author}{\bibfnamefont{L.~F.} \bibnamefont{Spietz}},
  \bibinfo{author}{\bibfnamefont{D.}~\bibnamefont{Gunnarsson}},
  \bibinfo{author}{\bibfnamefont{D.~I.} \bibnamefont{Schuster}},
  \bibinfo{author}{\bibfnamefont{P.}~\bibnamefont{Delsing}}, \bibnamefont{and}
  \bibinfo{author}{\bibfnamefont{R.~J.} \bibnamefont{Schoelkopf}},
  \bibinfo{journal}{Phys. Rev. Lett.} \textbf{\bibinfo{volume}{90}},
  \bibinfo{pages}{027002} (\bibinfo{year}{2003}).

\bibitem[{\citenamefont{Duty et~al.}(2004)\citenamefont{Duty, Gunnarsson,
  Bladh, and Delsing}}]{duty04}
\bibinfo{author}{\bibfnamefont{T.}~\bibnamefont{Duty}},
  \bibinfo{author}{\bibfnamefont{D.}~\bibnamefont{Gunnarsson}},
  \bibinfo{author}{\bibfnamefont{K.}~\bibnamefont{Bladh}}, \bibnamefont{and}
  \bibinfo{author}{\bibfnamefont{P.}~\bibnamefont{Delsing}},
  \bibinfo{journal}{Phys. Rev. B} \textbf{\bibinfo{volume}{69}},
  \bibinfo{pages}{140503} (\bibinfo{year}{2004}).

\bibitem[{\citenamefont{Schoelkopf et~al.}(1998)\citenamefont{Schoelkopf,
  Wahlgren, Kozhevnikov, Delsing, and Prober}}]{schoelkopf98}
\bibinfo{author}{\bibfnamefont{R.~J.} \bibnamefont{Schoelkopf}},
  \bibinfo{author}{\bibfnamefont{P.}~\bibnamefont{Wahlgren}},
  \bibinfo{author}{\bibfnamefont{A.~A.} \bibnamefont{Kozhevnikov}},
  \bibinfo{author}{\bibfnamefont{P.}~\bibnamefont{Delsing}}, \bibnamefont{and}
  \bibinfo{author}{\bibfnamefont{D.~E.} \bibnamefont{Prober}},
  \bibinfo{journal}{Science} \textbf{\bibinfo{volume}{280}},
  \bibinfo{pages}{1238} (\bibinfo{year}{1998}).

\bibitem[{\citenamefont{Cassidy et~al.}(2007)\citenamefont{Cassidy, Dzurak,
  Clark, Petersson, Farrer, Ritchie, and Smith}}]{cassidy07}
\bibinfo{author}{\bibfnamefont{M.~C.} \bibnamefont{Cassidy}},
  \bibinfo{author}{\bibfnamefont{A.~S.} \bibnamefont{Dzurak}},
  \bibinfo{author}{\bibfnamefont{R.~G.} \bibnamefont{Clark}},
  \bibinfo{author}{\bibfnamefont{K.~D.} \bibnamefont{Petersson}},
  \bibinfo{author}{\bibfnamefont{I.}~\bibnamefont{Farrer}},
  \bibinfo{author}{\bibfnamefont{D.~A.} \bibnamefont{Ritchie}},
  \bibnamefont{and} \bibinfo{author}{\bibfnamefont{C.~G.} \bibnamefont{Smith}},
  \bibinfo{journal}{App. Phys. Lett.} \textbf{\bibinfo{volume}{91}},
  \bibinfo{eid}{222104} (\bibinfo{year}{2007}).

\bibitem[{\citenamefont{Reilly et~al.}(2007)\citenamefont{Reilly, Marcus,
  Hanson, and Gossard}}]{reilly07}
\bibinfo{author}{\bibfnamefont{D.~J.} \bibnamefont{Reilly}},
  \bibinfo{author}{\bibfnamefont{C.~M.} \bibnamefont{Marcus}},
  \bibinfo{author}{\bibfnamefont{M.~P.} \bibnamefont{Hanson}},
  \bibnamefont{and} \bibinfo{author}{\bibfnamefont{A.~C.}
  \bibnamefont{Gossard}}, \bibinfo{journal}{App. Phys. Lett.}
  \textbf{\bibinfo{volume}{91}}, \bibinfo{eid}{162101} (\bibinfo{year}{2007}).

\bibitem[{\citenamefont{Wallraff et~al.}(2004)\citenamefont{Wallraff, Schuster,
  Blais, Frunzio, Huang, Majer, Kumar, Girvin, and Schoelkopf}}]{wallraff04}
\bibinfo{author}{\bibfnamefont{A.}~\bibnamefont{Wallraff}},
  \bibinfo{author}{\bibfnamefont{D.~I.} \bibnamefont{Schuster}},
  \bibinfo{author}{\bibfnamefont{A.}~\bibnamefont{Blais}},
  \bibinfo{author}{\bibfnamefont{L.}~\bibnamefont{Frunzio}},
  \bibinfo{author}{\bibfnamefont{R.-S.} \bibnamefont{Huang}},
  \bibinfo{author}{\bibfnamefont{J.}~\bibnamefont{Majer}},
  \bibinfo{author}{\bibfnamefont{S.}~\bibnamefont{Kumar}},
  \bibinfo{author}{\bibfnamefont{S.~M.} \bibnamefont{Girvin}},
  \bibnamefont{and} \bibinfo{author}{\bibfnamefont{R.~J.}
  \bibnamefont{Schoelkopf}}, \bibinfo{journal}{Nature}
  \textbf{\bibinfo{volume}{431}}, \bibinfo{pages}{162} (\bibinfo{year}{2004}).

\bibitem[{\citenamefont{Duty et~al.}(2005)\citenamefont{Duty, Johansson, Bladh,
  Gunnarsson, Wilson, and Delsing}}]{duty05}
\bibinfo{author}{\bibfnamefont{T.}~\bibnamefont{Duty}},
  \bibinfo{author}{\bibfnamefont{G.}~\bibnamefont{Johansson}},
  \bibinfo{author}{\bibfnamefont{K.}~\bibnamefont{Bladh}},
  \bibinfo{author}{\bibfnamefont{D.}~\bibnamefont{Gunnarsson}},
  \bibinfo{author}{\bibfnamefont{C.}~\bibnamefont{Wilson}}, \bibnamefont{and}
  \bibinfo{author}{\bibfnamefont{P.}~\bibnamefont{Delsing}},
  \bibinfo{journal}{Phys. Rev. Lett.} \textbf{\bibinfo{volume}{95}},
  \bibinfo{pages}{206807} (\bibinfo{year}{2005}).

\bibitem[{\citenamefont{Sillanp\"a\"a et~al.}(2005)\citenamefont{Sillanp\"a\"a,
  Lehtinen, Paila, Makhlin, Roschier, and Hakonen}}]{sillanpaa05}
\bibinfo{author}{\bibfnamefont{M.~A.} \bibnamefont{Sillanp\"a\"a}},
  \bibinfo{author}{\bibfnamefont{T.}~\bibnamefont{Lehtinen}},
  \bibinfo{author}{\bibfnamefont{A.}~\bibnamefont{Paila}},
  \bibinfo{author}{\bibfnamefont{Y.}~\bibnamefont{Makhlin}},
  \bibinfo{author}{\bibfnamefont{L.}~\bibnamefont{Roschier}}, \bibnamefont{and}
  \bibinfo{author}{\bibfnamefont{P.~J.} \bibnamefont{Hakonen}},
  \bibinfo{journal}{Phys. Rev. Lett.} \textbf{\bibinfo{volume}{95}},
  \bibinfo{pages}{206806} (\bibinfo{year}{2005}).

\bibitem[{\citenamefont{Blais et~al.}(2004)\citenamefont{Blais, Huang,
  Wallraff, Girvin, and Schoelkopf}}]{blais04}
\bibinfo{author}{\bibfnamefont{A.}~\bibnamefont{Blais}},
  \bibinfo{author}{\bibfnamefont{R.-S.} \bibnamefont{Huang}},
  \bibinfo{author}{\bibfnamefont{A.}~\bibnamefont{Wallraff}},
  \bibinfo{author}{\bibfnamefont{S.~M.} \bibnamefont{Girvin}},
  \bibnamefont{and} \bibinfo{author}{\bibfnamefont{R.~J.}
  \bibnamefont{Schoelkopf}}, \bibinfo{journal}{Phys. Rev. A}
  \textbf{\bibinfo{volume}{69}}, \bibinfo{eid}{062320} (\bibinfo{year}{2004}).

\bibitem[{pla()}]{plane}
\bibinfo{note}{For both the resonator and rf-QPC, the demodulated response is
  sampled and averaged while $V_L$ is ramped at 40 Hz. Planes are subtracted
  from the data to compensate for gate coupling and an offset applied to each
  vertical sweep to correct for low frequency drift in the demodulated
  response.}

\bibitem[{\citenamefont{Hayashi et~al.}(2003)\citenamefont{Hayashi, Fujisawa,
  Cheong, Jeong, and Hirayama}}]{hayashi03}
\bibinfo{author}{\bibfnamefont{T.}~\bibnamefont{Hayashi}},
  \bibinfo{author}{\bibfnamefont{T.}~\bibnamefont{Fujisawa}},
  \bibinfo{author}{\bibfnamefont{H.~D.} \bibnamefont{Cheong}},
  \bibinfo{author}{\bibfnamefont{Y.~H.} \bibnamefont{Jeong}}, \bibnamefont{and}
  \bibinfo{author}{\bibfnamefont{Y.}~\bibnamefont{Hirayama}},
  \bibinfo{journal}{Phys. Rev. Lett.} \textbf{\bibinfo{volume}{91}},
  \bibinfo{pages}{226804} (\bibinfo{year}{2003}).

\bibitem[{\citenamefont{Ota et~al.}(2009)\citenamefont{Ota, Hayashi, Muraki,
  and Fujisawa}}]{ota09}
\bibinfo{author}{\bibfnamefont{T.}~\bibnamefont{Ota}},
  \bibinfo{author}{\bibfnamefont{T.}~\bibnamefont{Hayashi}},
  \bibinfo{author}{\bibfnamefont{K.}~\bibnamefont{Muraki}}, \bibnamefont{and}
  \bibinfo{author}{\bibfnamefont{T.}~\bibnamefont{Fujisawa}},
  \bibinfo{journal}{arXiv:0910.2512}  (\bibinfo{year}{2009}).

\bibitem[{\citenamefont{DiCarlo et~al.}(2004)\citenamefont{DiCarlo, Lynch,
  Johnson, Childress, Crockett, Marcus, Hanson, and Gossard}}]{dicarlo04}
\bibinfo{author}{\bibfnamefont{L.}~\bibnamefont{DiCarlo}},
  \bibinfo{author}{\bibfnamefont{H.~J.} \bibnamefont{Lynch}},
  \bibinfo{author}{\bibfnamefont{A.~C.} \bibnamefont{Johnson}},
  \bibinfo{author}{\bibfnamefont{L.~I.} \bibnamefont{Childress}},
  \bibinfo{author}{\bibfnamefont{K.}~\bibnamefont{Crockett}},
  \bibinfo{author}{\bibfnamefont{C.~M.} \bibnamefont{Marcus}},
  \bibinfo{author}{\bibfnamefont{M.~P.} \bibnamefont{Hanson}},
  \bibnamefont{and} \bibinfo{author}{\bibfnamefont{A.~C.}
  \bibnamefont{Gossard}}, \bibinfo{journal}{Phys. Rev. Lett.}
  \textbf{\bibinfo{volume}{92}}, \bibinfo{pages}{226801}
  (\bibinfo{year}{2004}).

\bibitem[{\citenamefont{Buehler et~al.}(2004)\citenamefont{Buehler, Reilly,
  Starrett, Court, Hamilton, Dzurak, and Clark}}]{buehler04}
\bibinfo{author}{\bibfnamefont{T.~M.} \bibnamefont{Buehler}},
  \bibinfo{author}{\bibfnamefont{D.~J.} \bibnamefont{Reilly}},
  \bibinfo{author}{\bibfnamefont{R.~P.} \bibnamefont{Starrett}},
  \bibinfo{author}{\bibfnamefont{N.~A.} \bibnamefont{Court}},
  \bibinfo{author}{\bibfnamefont{A.~R.} \bibnamefont{Hamilton}},
  \bibinfo{author}{\bibfnamefont{A.~S.} \bibnamefont{Dzurak}},
  \bibnamefont{and} \bibinfo{author}{\bibfnamefont{R.~G.} \bibnamefont{Clark}},
  \bibinfo{journal}{J. App. Phys.} \textbf{\bibinfo{volume}{96}},
  \bibinfo{pages}{4508} (\bibinfo{year}{2004}).

\bibitem[{\citenamefont{Cheong et~al.}(2002)\citenamefont{Cheong, Fujisawa,
  Hayashi, Hirayama, and Jeong}}]{cheong02}
\bibinfo{author}{\bibfnamefont{H.~D.} \bibnamefont{Cheong}},
  \bibinfo{author}{\bibfnamefont{T.}~\bibnamefont{Fujisawa}},
  \bibinfo{author}{\bibfnamefont{T.}~\bibnamefont{Hayashi}},
  \bibinfo{author}{\bibfnamefont{Y.}~\bibnamefont{Hirayama}}, \bibnamefont{and}
  \bibinfo{author}{\bibfnamefont{Y.~H.} \bibnamefont{Jeong}},
  \bibinfo{journal}{App. Phys. Lett.} \textbf{\bibinfo{volume}{81}},
  \bibinfo{pages}{3257} (\bibinfo{year}{2002}).

\bibitem[{fie()}]{field}
\bibinfo{note}{We note that there was a small stray field present ($\sim 2.5$
  mT) which we did not correct for.}

\bibitem[{\citenamefont{Churchill et~al.}(2009)\citenamefont{Churchill,
  Kuemmeth, Harlow, Bestwick, Rashba, Flensberg, Stwertka, Taychatanapat,
  Watson, and Marcus}}]{churchill09}
\bibinfo{author}{\bibfnamefont{H.~O.~H.} \bibnamefont{Churchill}},
  \bibinfo{author}{\bibfnamefont{F.}~\bibnamefont{Kuemmeth}},
  \bibinfo{author}{\bibfnamefont{J.~W.} \bibnamefont{Harlow}},
  \bibinfo{author}{\bibfnamefont{A.~J.} \bibnamefont{Bestwick}},
  \bibinfo{author}{\bibfnamefont{E.~I.} \bibnamefont{Rashba}},
  \bibinfo{author}{\bibfnamefont{K.}~\bibnamefont{Flensberg}},
  \bibinfo{author}{\bibfnamefont{C.~H.} \bibnamefont{Stwertka}},
  \bibinfo{author}{\bibfnamefont{T.}~\bibnamefont{Taychatanapat}},
  \bibinfo{author}{\bibfnamefont{S.~K.} \bibnamefont{Watson}},
  \bibnamefont{and} \bibinfo{author}{\bibfnamefont{C.~M.}
  \bibnamefont{Marcus}}, \bibinfo{journal}{Phys. Rev. Lett.}
  \textbf{\bibinfo{volume}{102}}, \bibinfo{eid}{166802} (\bibinfo{year}{2009}).

\bibitem[{\citenamefont{Reilly et~al.}(2008)\citenamefont{Reilly, Taylor,
  Petta, Marcus, Hanson, and Gossard}}]{reilly08}
\bibinfo{author}{\bibfnamefont{D.~J.} \bibnamefont{Reilly}},
  \bibinfo{author}{\bibfnamefont{J.~M.} \bibnamefont{Taylor}},
   \bibinfo{author}{\bibfnamefont{E.~A.} \bibnamefont{Laird}},
  \bibinfo{author}{\bibfnamefont{J.~R.} \bibnamefont{Petta}},
  \bibinfo{author}{\bibfnamefont{C.~M.} \bibnamefont{Marcus}},
  \bibinfo{author}{\bibfnamefont{M.~P.} \bibnamefont{Hanson}},
  \bibnamefont{and} \bibinfo{author}{\bibfnamefont{A.~C.}
  \bibnamefont{Gossard}}, \bibinfo{journal}{Phys. Rev. Lett.}
  \textbf{\bibinfo{volume}{101}}, \bibinfo{pages}{236803} (\bibinfo{year}{2008}).

\bibitem[{\citenamefont{Johansson et~al.}(2006)\citenamefont{Johansson,
  Tornberg, and Wilson}}]{johansson06}
\bibinfo{author}{\bibfnamefont{G.}~\bibnamefont{Johansson}},
  \bibinfo{author}{\bibfnamefont{L.}~\bibnamefont{Tornberg}}, \bibnamefont{and}
  \bibinfo{author}{\bibfnamefont{C.~M.} \bibnamefont{Wilson}},
  \bibinfo{journal}{Phys. Rev. B} \textbf{\bibinfo{volume}{74}},
  \bibinfo{eid}{100504} (\bibinfo{year}{2006}).

\bibitem[{\citenamefont{Andresen et~al.}(2008)\citenamefont{Andresen, Wu,
  Danneau, Gunnarsson, and Hakonen}}]{andresen08}
\bibinfo{author}{\bibfnamefont{S.~E.~S.} \bibnamefont{Andresen}},
  \bibinfo{author}{\bibfnamefont{F.}~\bibnamefont{Wu}},
  \bibinfo{author}{\bibfnamefont{R.}~\bibnamefont{Danneau}},
  \bibinfo{author}{\bibfnamefont{D.}~\bibnamefont{Gunnarsson}},
  \bibnamefont{and} \bibinfo{author}{\bibfnamefont{P.~J.}
  \bibnamefont{Hakonen}}, \bibinfo{journal}{J. App. Phys.}
  \textbf{\bibinfo{volume}{104}}, \bibinfo{eid}{033715} (\bibinfo{year}{2008}).

\bibitem[{\citenamefont{Gustavsson et~al.}(2007)\citenamefont{Gustavsson,
  Studer, Leturcq, Ihn, Ensslin, Driscoll, and Gossard}}]{gustavsson07}
\bibinfo{author}{\bibfnamefont{S.}~\bibnamefont{Gustavsson}},
  \bibinfo{author}{\bibfnamefont{M.}~\bibnamefont{Studer}},
  \bibinfo{author}{\bibfnamefont{R.}~\bibnamefont{Leturcq}},
  \bibinfo{author}{\bibfnamefont{T.}~\bibnamefont{Ihn}},
  \bibinfo{author}{\bibfnamefont{K.}~\bibnamefont{Ensslin}},
  \bibinfo{author}{\bibfnamefont{D.~C.} \bibnamefont{Driscoll}},
  \bibnamefont{and} \bibinfo{author}{\bibfnamefont{A.~C.}
  \bibnamefont{Gossard}}, \bibinfo{journal}{Phys. Rev. Lett.}
  \textbf{\bibinfo{volume}{99}}, \bibinfo{eid}{206804} (\bibinfo{year}{2007}).

\bibitem[{\citenamefont{Petersson et~al.}(2009)\citenamefont{Petersson, Smith,
  Anderson, Atkinson, Jones, and Ritchie}}]{petersson09}
\bibinfo{author}{\bibfnamefont{K.~D.} \bibnamefont{Petersson}},
  \bibinfo{author}{\bibfnamefont{C.~G.} \bibnamefont{Smith}},
  \bibinfo{author}{\bibfnamefont{D.}~\bibnamefont{Anderson}},
  \bibinfo{author}{\bibfnamefont{P.}~\bibnamefont{Atkinson}},
  \bibinfo{author}{\bibfnamefont{G.~A.~C.} \bibnamefont{Jones}},
  \bibnamefont{and} \bibinfo{author}{\bibfnamefont{D.~A.}
  \bibnamefont{Ritchie}}, \bibinfo{journal}{Phys. Rev. Lett.}
  \textbf{\bibinfo{volume}{103}}, \bibinfo{pages}{016805}
  (\bibinfo{year}{2009}).

\bibitem[{\citenamefont{Hu et~al.}(2007)\citenamefont{Hu, Churchill, Reilly,
  Xiang, Lieber, and Marcus}}]{hu07}
\bibinfo{author}{\bibfnamefont{Y.}~\bibnamefont{Hu}},
  \bibinfo{author}{\bibfnamefont{H.~O.~H.} \bibnamefont{Churchill}},
  \bibinfo{author}{\bibfnamefont{D.~J.} \bibnamefont{Reilly}},
  \bibinfo{author}{\bibfnamefont{J.}~\bibnamefont{Xiang}},
  \bibinfo{author}{\bibfnamefont{C.~M.} \bibnamefont{Lieber}},
  \bibnamefont{and} \bibinfo{author}{\bibfnamefont{C.~M.}
  \bibnamefont{Marcus}}, \bibinfo{journal}{Nat. Nano.}
  \textbf{\bibinfo{volume}{2}}, \bibinfo{pages}{622} (\bibinfo{year}{2007}).

\bibitem[{\citenamefont{Biercuk et~al.}(2006)\citenamefont{Biercuk, Reilly,
  Buehler, Chan, Chow, Clark, and Marcus}}]{biercuk06}
\bibinfo{author}{\bibfnamefont{M.~J.} \bibnamefont{Biercuk}},
  \bibinfo{author}{\bibfnamefont{D.~J.} \bibnamefont{Reilly}},
  \bibinfo{author}{\bibfnamefont{T.~M.} \bibnamefont{Buehler}},
  \bibinfo{author}{\bibfnamefont{V.~C.} \bibnamefont{Chan}},
  \bibinfo{author}{\bibfnamefont{J.~M.} \bibnamefont{Chow}},
  \bibinfo{author}{\bibfnamefont{R.~G.} \bibnamefont{Clark}}, \bibnamefont{and}
  \bibinfo{author}{\bibfnamefont{C.~M.} \bibnamefont{Marcus}},
  \bibinfo{journal}{Phys. Rev. B} \textbf{\bibinfo{volume}{73}},
  \bibinfo{eid}{201402} (\bibinfo{year}{2006}).

\end{thebibliography}
\end{document}